\documentclass{PoS}

\usepackage{psfig}
\usepackage{graphicx}
\usepackage{graphics}

\title{Longitudinal flow and onset of deconfinement}

\ShortTitle{Longitudinal flow and onset of deconfinement}

\author{Hannah Petersen
	\thanks{H.P. gratefully acknowledges financial support by a stipendship of the Deutsche Telekom-Stiftung}\\

        Institut f\"ur Theoretische Physik, Johann Wolfgang Goethe-Universit\"at, Max-von-Laue-Str.~1, D-60438 Frankfurt am Main, Germany\\
        E-mail: \email{petersen@th.physik.uni-frankfurt.de}}

\author{\speaker{Marcus Bleicher}\\
         
        Institut f\"ur Theoretische Physik, Johann Wolfgang Goethe-Universit\"at, Max-von-Laue-Str.~1, D-60438 Frankfurt am Main, Germany\\
        E-mail: \email{bleicher@th.physik.uni-frankfurt.de}}

\abstract{The effects of the onset of deconfinement on longitudinal and transverse flow are studied. First, we analyze longitudinal pion spectra from $E_{\rm lab}= 2A$~GeV to $\sqrt{s_{\rm NN}}=200$~GeV within Landau's hydrodynamical model and the UrQMD transport approach. From the measured data on the widths of the pion rapidity spectra, we extract the sound velocity $c_s^2$ in the early stage of the reactions. It is found that the sound velocity has a local minimum (indicating a softest point in the equation of state, EoS) at $E_{\rm beam}=30A$~GeV. This softening of the EoS is compatible with the assumption of the formation of a mixed phase at the onset of deconfinement. Furthermore, the energy excitation function of elliptic flow ($v_2$) from $E_{\rm beam}=90A$~MeV to $\sqrt{s_{\rm NN}}=200$~GeV is explored within the UrQMD framework and discussed in the context of the available data. The transverse flow should also be sensitive to changes in the equation of state. Therefore, the underestimation of elliptic flow by the UrQMD model calculation above $E_{\rm lab}= 30A$~GeV might also be explained by assuming a phase transition from a hadron gas to the quark gluon plasma around this energy. This would be consistent with the model calculations, indicating a transition from hadronic matter to ``string matter'' in this energy range.}

\FullConference{Critical Point and Onset of Deconfinement\\
		 July 3-6 2006\\
		 Florence, Italy}

\begin{document}

\section{Introduction}
Over the last years, a wealth of detailed data in the $20A-160A$~GeV 
energy regime has become available. 
The systematic study of these data revealed surprising (non-monotonous) 
structures in various observables around $30A$~GeV beam energy.
Most notable irregular structures in that energy regime include, 
\begin{itemize}
\item 
the sharp maximum in the K$^+/\pi^+$ ratio \cite{Afanasiev:2002mx,Gazdzicki:2004ef},
\item
a step in the transverse momentum excitation function (as seen through 
$\langle m_\perp\rangle -m_0$ ) \cite{Gazdzicki:2004ef,na49_blume},
\item
an apparent change in the pion per participant ratio \cite{Gazdzicki:2004ef} and
\item
increased ratio fluctuations (due to missing data at low energies it is unknown if this 
is a local maximum or an ongoing increase of the fluctuations) \cite{Roland:2005pr}.
\end{itemize}

It has been speculated, that these observation hint towards the onset of deconfinement
already at $30A$~GeV beam energy. Indeed, increased strangeness production \cite{Koch:1986ud} 
and enhanced fluctuations have long been predicted as a sign of QGP 
formation \cite{Bleicher:2000ek,Jeon:2005kj,Shuryak:2000pd,Heiselberg:2000ti,Muller:2001wj,Gazdzicki:2003bb,Gorenstein:2003hk} within different 
frameworks and observables.
The suggestion of an enhanced strangeness to entropy ratio ($\sim K/\pi$) as indicator for the onset of QGP formation 
was especially advocated in \cite{SMES}. Also  the  high and approximately
constant $K^\pm$ inverse slopes of the $m_T$ spectra above $\sim 30A$~GeV - the 'step' - was also found to be consistent
with the assumption of a parton $\leftrightarrow$ hadron phase transition at low SPS 
energies \cite{Gorenstein:2003cu,Hama:2004re}.  
Surprisingly, transport simulations (supplemented by recent lattice QCD (lQCD) calculations) 
have also suggested that partonic degrees of freedom might already lead to
visible effects at $\sim 30A$~GeV \cite{Weber98,MT-prl,Bratkovskaya:2004kv}. 
Finally, the comparison of the thermodynamic parameters $T$ and $\mu_B$
extracted from the transport models in the central overlap region
\cite{Bravina} with the experimental systematics on chemical
freeze-out configurations \cite{Braun-Munzinger:1996mq,Braun-Munzinger:1998cg,Cleymans} 
in the $T-\mu_B$ plane do also suggest that a first glimpse on a deconfined state might be possible
around $10A-30A$~GeV.

In the first part of this paper, we explore wether similar irregularities are also present in the excitation function of longitudinal observables, namely rapidity distributions. Here we employ Landau's hydrodynamical model and the UrQMD transport approach. In the second part, we focus on the excitation function of transverse flow ($v_2$) and discuss UrQMD results in the context of the available data.    

\section{Longitudinal flow}

It became popular
to interpret relativistic heavy ion reactions with Landau's
hydrodynamical model
\cite{Fermi:1950jd,Landau:gs,Belenkij:cd,Carruthers:ws,Carruthers:dw,Carruthers:1981vs}
(for recent applications of this model to relativistic
nucleus-nucleus interactions see
\cite{Stachel:1989pa,Steinberg:2004vy,Murray:2004gh,Roland:2004,Bleicher:2005ys,Bleicher:2005tb}). Therefore we
will use this simple hydrodynamical picture as a baseline for the
model and data comparison. The main physics assumptions of Landau's
picture are: The collision of two Lorentz-contracted hadrons or
nuclei leads to full thermalization in a volume of size
$V{m_p}/\sqrt{s}$. This justifies the use of thermodynamics and
establishes the system size and energy dependence. Usually a simple equation
of state  $p=\epsilon/3$ is assumed. Chemical potentials are usually
assumed to vanish. The main results derived from these assumptions
are: A universal formula for the produced entropy, determined mainly
by the initial Lorentz contraction and Gaussian rapidity distributions,
at least for newly produced particles. Under the condition that $c_s$ is independent of temperature, 
the rapidity density is given by \cite{Carruthers:dw,Shuryak:1972zq}:
\begin{equation}
\frac{dN}{dy}=\frac{Ks_{\rm NN}^{1/4}}{\sqrt{2\pi \sigma_y^2}}\,\exp\left(-\frac{y^2}{2\sigma_y^2}\right)
\label{eq1}
\end{equation}
with
\begin{equation}
\sigma_y^2=\frac{8}{3}\frac{c_s^2}{1-c_s^4}\,{\rm ln}({\sqrt {s_{\rm NN}}}/{2m_p})\quad,
\label{eq2}
\end{equation}
where $K$ is a normalisation factor and $m_p$ is the proton mass.

\begin{figure}
\begin{tabular*}{16cm}{ll}
  \psfig{file=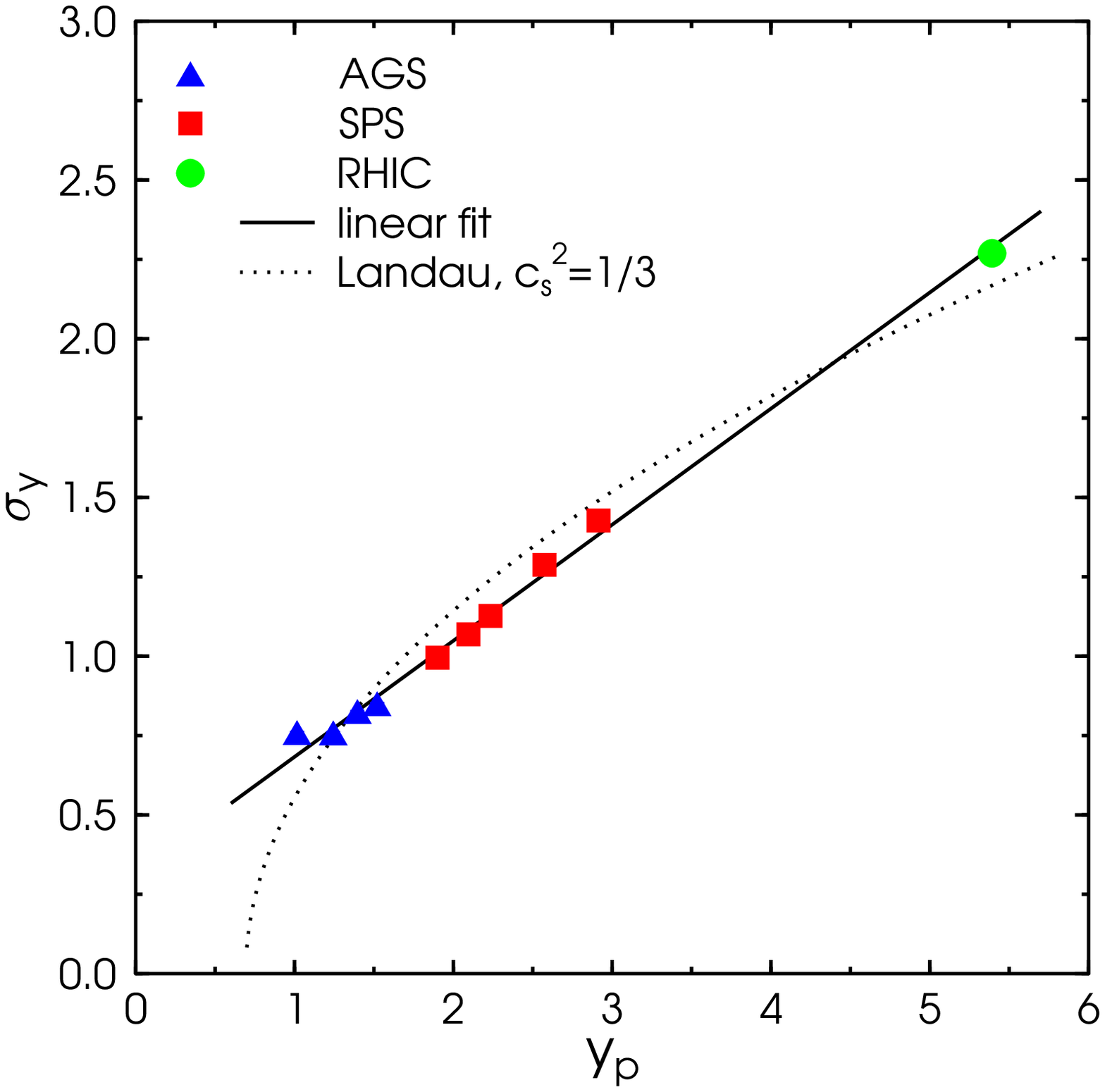,width=7.5cm} & \psfig{file=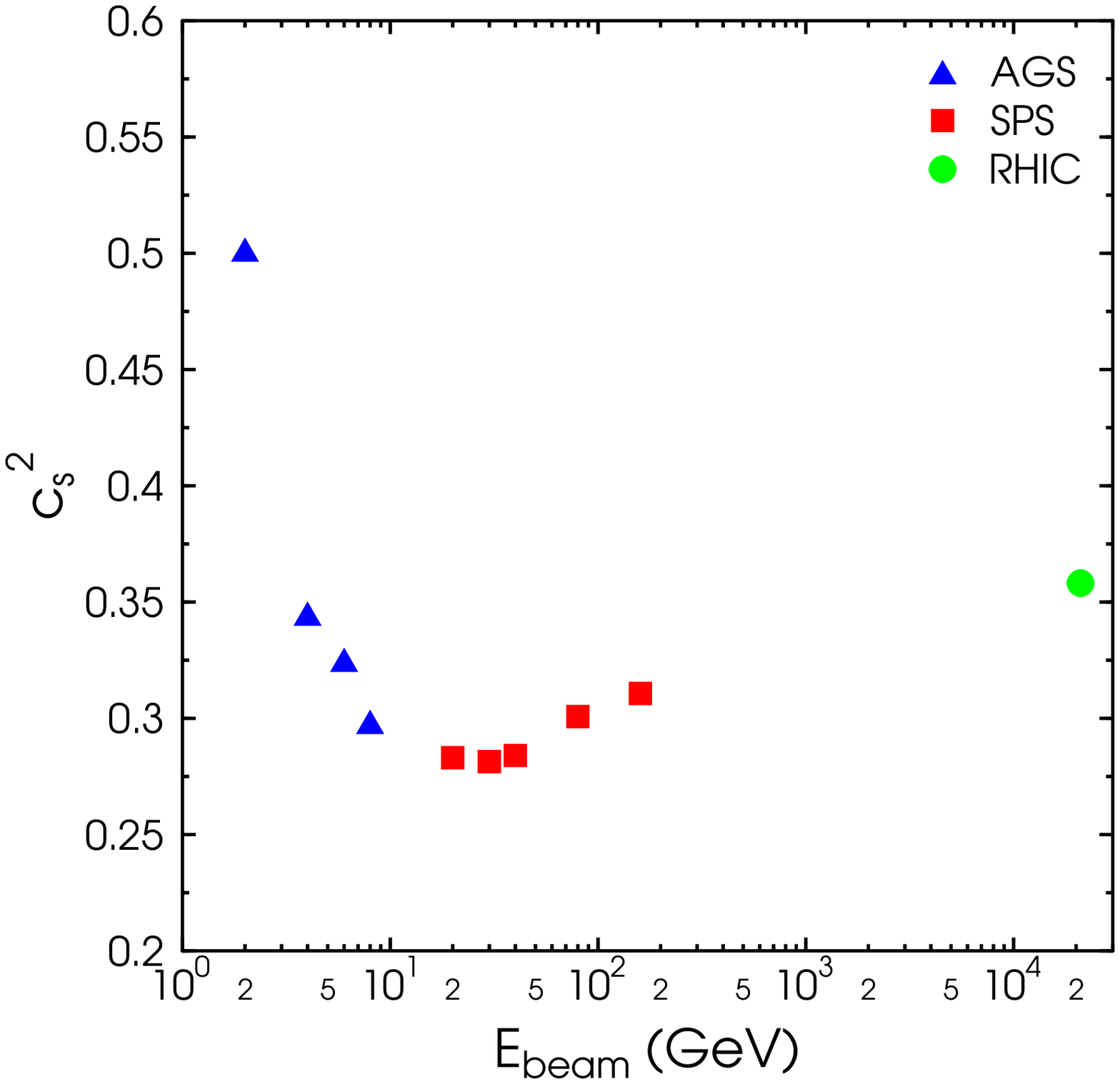,width=7.5cm} \\
\end{tabular*}\vspace*{-0.3cm}
\caption{Left: \protect\label{rapwidth}The root mean square $\sigma_y$ of the rapidity
distributions of negatively charged pions in central Pb+Pb (Au+Au) reactions 
as a function of the beam rapidity $y_p$.
The dotted line indicates the Landau model prediction with  $c_s^2=1/3$, while
the full line shows a linear fit through the data points.
Data (full symbols) are taken from \protect\cite{na49_blume,Roland:2004,klay,brahms}.
The statistical errors given by the experiments are smaller than the symbol sizes. Systematic errors are not available. Right: \protect\label{c02} Speed of sound as a function of beam energy for central 
Pb+Pb (Au+Au) reactions as extracted from the data using Eq.\ (\protect\ref{eq3}).
The statistical errors (not shown) are smaller than 3\%.}
\end{figure}

Let us now analyze the available experimental data on rapidity distributions of negatively 
charged pions in terms of the Landau model.
Fig. \ref{rapwidth} (left) shows the measured root mean square $\sigma_y$ of the rapidity
distribution of negatively charged pions in central Pb+Pb (Au+Au) reactions 
as a function of the beam rapidity. The dotted line indicates 
the Landau model predictions with  the commonly used constant sound velocity $c_s^2=1/3$. 
The full line shows a linear fit through the data points, while the
data points \cite{na49_blume,Roland:2004,klay,brahms} are depicted by full symbols.

At a first glance the energy dependence looks structureless.
The data seem to follow a linear dependence on the beam rapidity $y_p$ without
any irregularities.
However, the general trend of the rapidity widths is also well reproduced by 
Landau's model with an equation of state with a fixed speed of sound. 
Nevertheless, there seem to be systematic deviations.
At low AGS energies and at RHIC, the experimental points are generally
underpredicted by Eq.\ (\ref{eq2}), while in the SPS energy regime Landau's model overpredicts the
widths of the rapidity distributions.
Exactly these deviations from the simple Landau picture do allow to 
gain information on the equation of state 
of the matter produced in the early stage of the reaction.
By inverting Eq.\ (\ref{eq2}) we can express the speed of sound $c_s^2$ in the medium as a function of 
the measured width of the rapidity distribution:
\begin{equation}
c_s^2=-\frac{4}{3}\frac{{\rm ln}({\sqrt {s_{\rm NN}}}/{2 m_p})}{\sigma_y^2}
+\sqrt{\left[\frac{4}{3}\frac{{\rm ln}({\sqrt {s_{\rm NN}}}/{2 m_p})}{\sigma_y^2}\right]^2+1}\quad.
\label{eq3}
\end{equation}

Let us now investigate the energy dependence of the sound velocities extracted
from the data. Fig. \ref{c02} (right) shows the speed of sound as a function of beam energy for central 
Pb+Pb (Au+Au) reactions as obtained from the data using Eq.\ (\ref{eq3}).
The sound velocities exhibit a clear  minimum (usually called the softest point) around a beam energy of
$30A$~GeV.
A localized softening of the equation of state is a long predicted  signal for the mixed phase 
at the transition energy from hadronic to partonic matter \cite{Hung:1994eq,Rischke:1995pe,Brachmann:1999mp}. 
\begin{figure}[t]
\begin{minipage} [r] {6cm}
\psfig{file=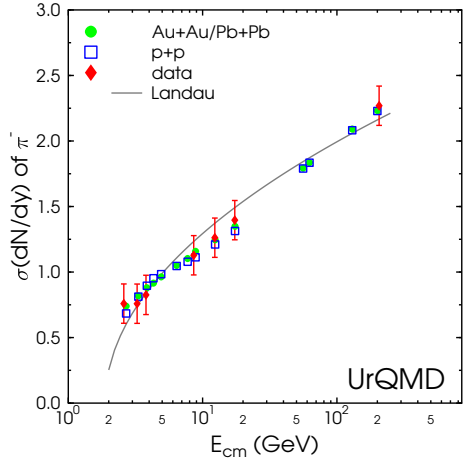,width=7.5cm}
\end{minipage}
\hspace*{-0.5cm}\begin{minipage} [r] {7.5cm}
\caption{\protect\label{urqmdrapwidth} The root mean square of the rapidity
distribution of negatively charged Pions in central Au+Au/Pb+Pb and
Proton+Proton reactions as a function of the center of mass energy.
UrQMD calculations for Au+Au/Pb+Pb are denoted by full circles, the
pp results are shown by open squares. The prediction from Landau's
model is given by the line (Eq. \protect\ref{eq1}). Data \protect\cite{Roland:2004}
are depicted by full diamonds.}
\end{minipage}

\end{figure}

To test this hypothesis the same observable has also been calculated using the UrQMD model (v2.2) \cite{UrQMD1,UrQMD2}. This transport model takes into
account the formation and multiple rescattering of hadrons and
dynamically describes the generation of pressure in the hadronic
expansion phase. It involves also interactions of (di-)quarks, however  
gluonic degrees of freedom are not treated explicitly, but are implicitly present  in strings. This simplified treatment is generally accepted to describe Proton-Proton and Proton-nucleus
interactions. 

As depicted in Fig. \ref{urqmdrapwidth} the UrQMD predictions (full circles) for the rapidity widths of negatively charged pions in Au+Au (Pb+Pb) reactions are in line with the experimental data \cite{Roland:2004} (full diamonds) 
and  Landau's hydrodynamical model (full line). A rather surprising observation is that the calculated rapidity widths of $\pi^-$ in pp interactions (open squares) are identical to the AA results. 
Together with the previous discussion, it seems that the equation of state in the transport model is also soft in the SPS regime. Thus, the nature of the softest point remains unclear. 

\section{Transverse flow}

Let us now look at the dynamics of the system perpendicular to the beam direction. The transverse flow is intimately connected to the pressure gradients. Therefore, it is sensitive to the equation of state (EoS) and might be used to search for abnormal matter states and phase transitions \cite{Stoecker:1979mj,Hofmann:1976dy,Stoecker:1986ci}. Especially the second coefficient of the Fourier expansion of the azimuthal distribution of the emitted particles ($v_2$) is a valuable tool to gain insight into the expanding stage of the fireball. 

\begin{figure}
\begin{tabular*}{16cm}{ll}
  \psfig{file=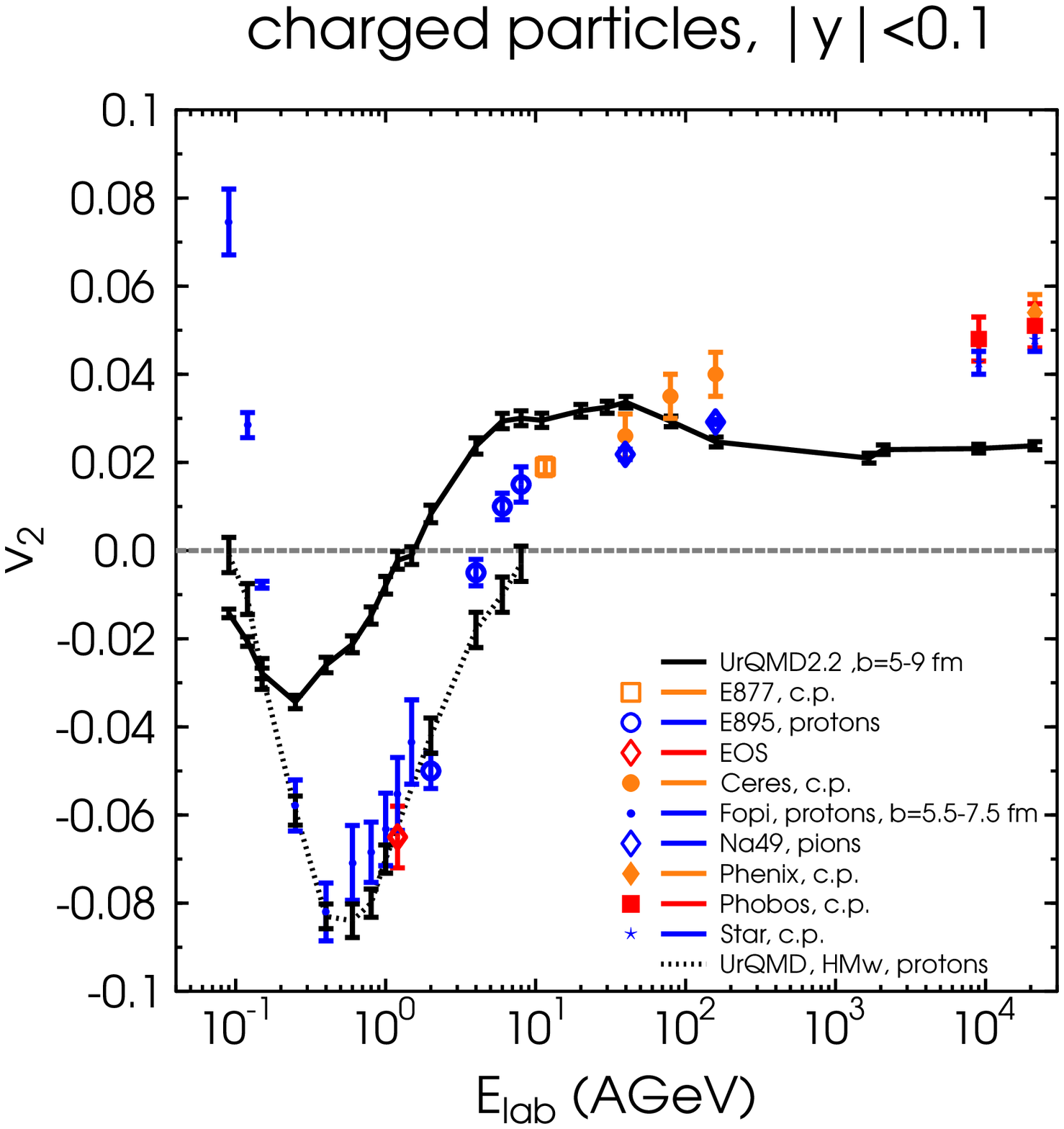,width=7.5cm} & \psfig{file=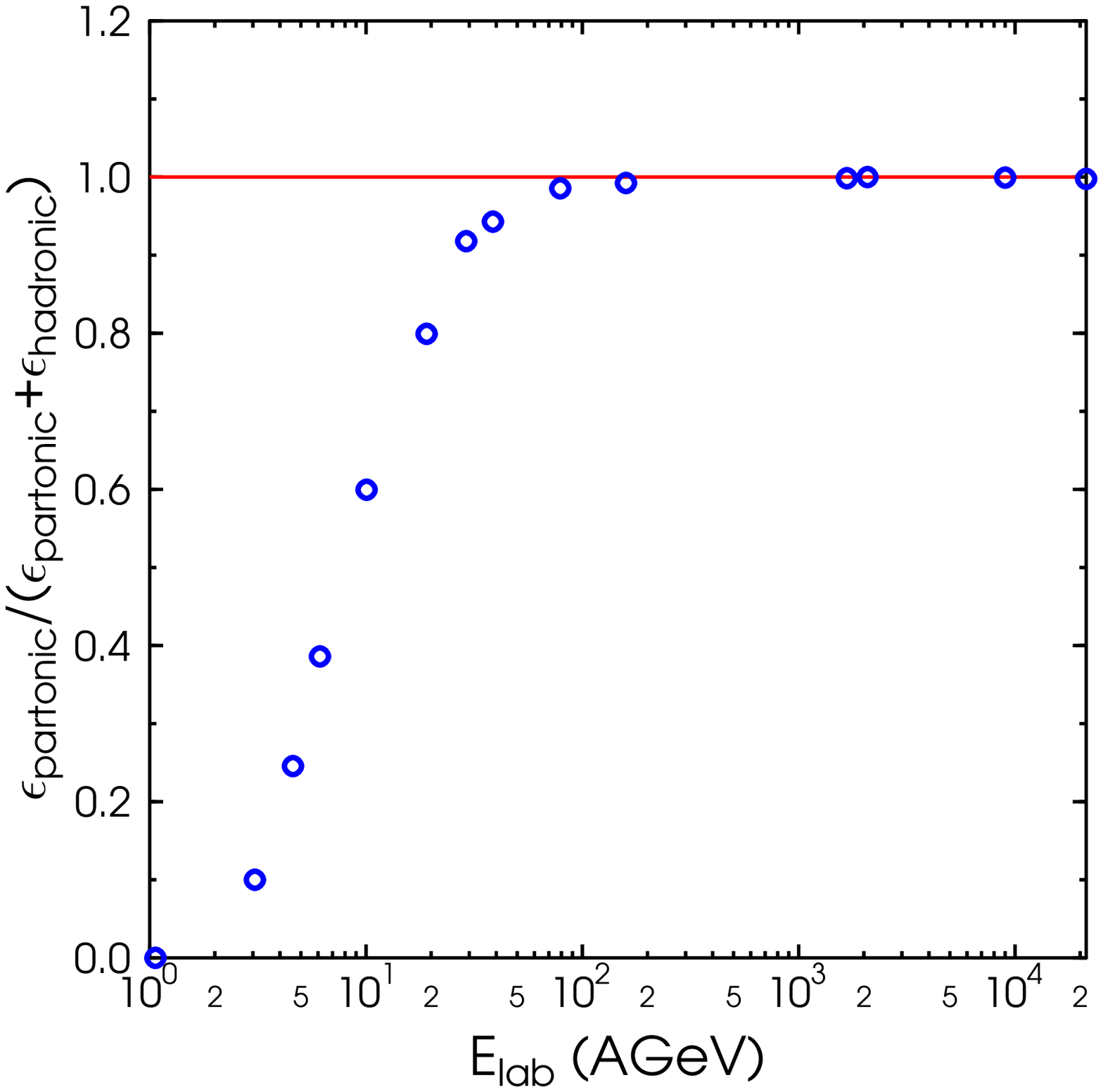,width=7.5cm} \\
\end{tabular*}\vspace*{-0.3cm}
\caption{Left: \label{figv2chexc}The calculated energy excitation function of elliptic flow of charged particles in Au+Au/Pb+Pb 
collisions in mid-central collisions (b=5-9 fm)with $|y|<0.1$(black line). This curve is compared to data from 
different experiments for mid-central collisions. For E895 \cite{Pinkenburg:1999nv,Chung:2001qr} and 
FOPI \cite{Andronic:2004cp} there is the elliptic flow of protons and for NA49 \cite{Alt:2003ab} it is the elliptic 
flow of pions. For E877, CERES \cite{Filimonov:2001fd,Slivova:2002wj,CERES:YUN02}, PHENIX \cite{Esumi:2002vy},
PHOBOS \cite{Manly:2002uq} and STAR \cite{Ray:2002md} there is data for the charged particle flow. The dotted line 
in the low energy regime depict UrQMD calculations with the mean field \cite{Li:2006ez}. Right:\label{fig_ed} Calculated fraction of energy density in unformed hadrons with $|y|<0.5$ and in a cylindrical volume with transverse radius $r=3$ fm and length $h=3/\gamma_{CM}$ fm as a function of the beam energy for central Pb+Pb (Au+Au) reactions.}
\end{figure}

The excitation function of charged particle elliptic flow is
compared to data over a wide energy range (Fig. \ref{figv2chexc} (left)), i.e from $E_{\rm beam}=90A$~MeV to 
$\sqrt{s_{NN}}=200$~GeV. The squeeze-out effect at low energies and the change to in-plane emission at higher 
energies is nicely observed in the excitation function. The symbols indicate the data for charged particles 
from different experiments. Note however, that in the low energy regime
there are only experimental data points for protons. For beam energies below
2A GeV most of the charged particles are also protons because there is not
enough energy to produce many new particles. Going to higher energies the
elliptic flow of pions and charged particles are very similar. The
rapidity cut of $|y| < 0.1$ has been used for the whole energy range
despite the fact that the data at higher energies is within $|y| < 0.5$. This has been done
to avoid too much changes in the parameters and this choice gives reasonable results over the whole energy range. 
We have checked that the results at higher energies are not affected by the choice of this narrower rapidity window.

At low energies $E_{\rm beam}\sim 0.1 - 6$A GeV the
squeeze-out effect, i.e. the elliptic flow out-of-plane, is clearly seen
in the data as well as in the calculations, especially when the mean field is considered. At such energies, it is well 
known that both the mean field and the two-body collision are equally important  to reproduce quantitatively the 
experimental results \cite{Danielewicz:1999vh,Danielewicz:1998vz,Pan:1992ef}. In this paper we adopt a hard equation 
of state  with momentum dependence (HMw) which 
was updated recently in the UrQMD model \cite{Li:2005gf,Li:2006ez}. 

In the SPS regime the model calculations are quite in line with
the data, especially with the NA49 results. For a more detailed discussion of directed and elliptic flow results from UrQMD-2.2 the reader is referred to \cite{Petersen:2006vm,Zhu:2006fb,Zhu:2005qa}. Above $E_{\rm lab}=160A~$GeV the calculation underestimates the elliptic 
flow. At the highest RHIC energy there are about 5\% flow in the data while the
model calculation provides only half of this value. This can be explained by
assuming a lack of pressure in the transport model at these energies.

It is possible that above the energy range about $E_{\rm lab}=30~A$GeV partonic interactions have to be
taken into account to describe the data as suggested in \cite{Bratkovskaya:2004kv,Weber:1998zb,Zhu:2006qg}. How can we analyse this
question in the model, since there are no partonic degrees of freedom explicitly incorporated?  In the current
model exists a formation time for hadrons produced in the string fragmentation. The leading hadrons of the
fragmenting strings contain the valence quarks of the original excited hadron.  These (di-)quark string ends are
allowed to interact during their formation time with a reduced cross section defined by the additive quark
model. Other ``pre-hadrons'' from the fragmenting string are not allowed to interact before the coalescence of the produced quarks. Thus, because the unformed hadrons do not interact with others during their formation time, the effective pressure is reduced and 
only build up from the density of the formed hadrons. 

To illuminate this, we have  calculated
the energy density during heavy ion collisions at different beam energies. From this, we
extract the time corresponding to the maximum value of the total energy density. Fig. \ref{fig_ed} (right) shows the fraction of the energy density that is deposited in
the ''unformed hadrons''($\epsilon_{partonic}/(\epsilon_{partonic}+\epsilon_{hadronic})$). I.e. all string fragments within their formation time are dubbed as ``partonic''. The fraction of $\epsilon_{partonic}$ starts at zero for low energies and then rises
fast to almost 100 $\%$. Note that this fraction reaches 90 $\%$ already around  $30$~AGeV beam energy, similar to
the energy region where a phase transition is expected. As one can see, the energy density of the formed hadrons ($\epsilon_{hadronic}$)
is much smaller than the total value, therefore the effective pressure of the formed hadrons alone in the model seems to be too small to generate enough $v_2$. Thus, this finding supports the interpretation
of the need for initial pressure from non-hadronic matter already at low SPS energies.
  
\section{Conclusion}

In conclusion, we have explored the excitation functions of the rapidity widths and of elliptic flow of (negatively) charged
pions in Pb+Pb (Au+Au) collisions. The following observations can be made:

\begin{itemize}
\item
The rapidity spectra of pions produced in  central nucleus-nucleus reactions at all investigated energies can be 
well described by single Gaussians.
\item
The energy dependence of the width of the pion rapidity distribution follows the
prediction of  Landau's hydrodynamical model if a variation of the sound
velocity  is taken into account.
\item
The speed of sound excitation function extracted from the data has a pronounced 
minimum (softest point) at $E_{\rm beam}=30A$~GeV.
\item 
The UrQMD model describes the rapidity widths data well, but underestimates the elliptic flow from the higher SPS energy on. 
\item 
The softest point coincides with the rapid rise of ``partonic'' degrees of freedom in the present model. 
\item
This softest point might be due to  the formation of a mixed phase indicating the onset of deconfinement at this energy.

\end{itemize}

Further explorations of this energy domain is needed and can be done at the future FAIR facility and 
by CERN-SPS and BNL-RHIC experiments.

\section*{Acknowledgements}

This work was supported by GSI and BMBF.
This work used computational resources provided by the
Center for Scientific Computing at Frankfurt (CSC).

\end{document}